\documentclass[sn-mathphys,Numbered]{sn-jnl}

\usepackage{graphicx}%
\usepackage{multirow}%
\usepackage{amsmath,amssymb,amsfonts}%
\usepackage{amsthm}%
\usepackage{mathrsfs}%
\usepackage[title]{appendix}%
\usepackage{xcolor}%
\usepackage{textcomp}%
\usepackage{manyfoot}%
\usepackage{booktabs}%
\usepackage{algorithm}%
\usepackage{algorithmicx}%
\usepackage{algpseudocode}%
\usepackage{listings}%
\usepackage{url}

\theoremstyle{thmstyleone}%
\theoremstyle{thmstyletwo}%

\theoremstyle{thmstylethree}%

\begin{document}

\title[Article Title]{Muon spin relaxation in mixed perovskite 
(LaAlO$_3$)$_{x}$(SrAl$_{0.5}$Ta$_{0.5}$O$_3$)$_{1-x}$ with $x\simeq 0.3$}


\author*[1]{\fnm{Takashi U.} \sur{Ito}}\email{tuito@post.j-parc.jp}

\author[1,2]{\fnm{Wataru} \sur{Higemoto}}

\author[3]{\fnm{Akihiro} \sur{Koda}}

\author[3]{\fnm{Jumpei G.} \sur{Nakamura}}

\author[3]{\fnm{Koichiro} \sur{Shimomura}}

\affil*[1]{\orgdiv{Advanced Science Research Center}, \orgname{Japan
Atomic Energy Agency}, \orgaddress{\street{Tokai}, \city{Ibaraki}, \postcode{319-1195}, \country{Japan}}}

\affil[2]{\orgdiv{Department of Physics}, \orgname{Tokyo Institute of
Technology}, \orgaddress{\street{Meguro}, \city{Tokyo}, \postcode{152-8551}, \country{Japan}}}

\affil[3]{\orgdiv{Institute of Materials Structure Science},
\orgname{High Energy Accelerator Research Organization (KEK)}, \orgaddress{\street{Tsukuba}, \city{Ibaraki}, \postcode{305-0801}, \country{Japan}}}


\abstract{We report on muon spin relaxation ($\mu^+$SR) measurements in 
a mixed perovskite compound, 
(LaAlO$_3$)$_{x}$(SrAl$_{0.5}$Ta$_{0.5}$O$_3$)$_{1-x}$ with $x\simeq 0.3$
(LSAT), which is widely used as a single-crystalline substrate for
thin film deposition. In zero applied field (ZF), muon depolarization due to
the distribution of nuclear dipole fields was observed in the
temperature range from 4~K to 270~K.
Interestingly, $\mu^+$SR time spectra in ZF maintained a
Gaussian-like feature over the entire range, while the
depolarization rate exhibited a monotonic decrease with increasing 
temperature. This behavior may be attributed to the thermally
activated diffusion of muons between a few adjacent sites within a 
confined space of the angstrom scale, where the motionally averaged
local field that each muon experiences can remain non-zero and result in
maintaining the Gaussian-like line shape.
The spatial distribution of electrostatic potential at lattice
interstices evaluated via density functional theory calculations suggests
that such a restriction of muon diffusion paths can be caused by the
random distribution of cations with different nominal valences in the
mixed perovskite lattice.}




\maketitle

\section{Introduction}\label{sec1}
The ultra-slow muon beamline (U-line), currently under commissioning at
 the Materials and Life Science Experimental Facility in the Japan
 Proton Accelerator Research Complex (J-PARC), enables positive muons
 ($\mu^+$) to be implanted into thin-film materials for $\mu^+$ spin rotation and relaxation ($\mu^+$SR) spectroscopy with nanometer depth resolution~\cite{kanda23}.
 The $\mu^+$SR spectrometer for the U-line is in the final
 stages of tune-ups and is expected to become available to public users
 shortly.
 
 In anticipation of forthcoming $\mu^+$SR experiments using ultra-slow muons,
we have investigated $\mu^+$SR signals from typical substrate materials, 
which can be superposed on those from thin-film samples deposited on the
substrates.
In this paper, we specifically focus on a mixed perovskite substrate
 material, (LaAlO$_3$)$_{x}$(SrAl$_{0.5}$Ta$_{0.5}$O$_3$)$_{1-x}$ with
 $x\simeq 0.3$ (LSAT)~\cite{mateika91}.
 LSAT is obtained as a solid solution of LaAlO$_3$ and
 SrAl$_{0.5}$Ta$_{0.5}$O$_3$ and has a cubic $AB$O$_3$ perovskite
 structure at the molar ratio of $\sim$3:7. The $A$ and $B$ cation sites are
 statistically occupied by Sr$^{2+}$ and La$^{3+}$, and Al$^{3+}$ and
 Ta$^{5+}$, respectively. Single-crystalline substrates of LSAT were 
 developed to overcome difficulties found in LaAlO$_3$ substrates, such
 as twinning, strain, and non-isotropic microwave
 properties~\cite{tidrow97}, and have been widely used for the epitaxial
 growth of various thin films, primarily of perovskite-related
 materials~\cite{tidrow97,kumar08,krockenberger10,verma15,goian17} and GaN semiconductors~\cite{sumiya02}.

  In this paper, we report $\mu^+$SR measurements on LSAT in zero
 applied field (ZF) below room temperature. We also present the results of
 density functional theory (DFT) calculations for an approximate model
 of LSAT, which were performed to gain deeper insight into 
 the localization site and kinetics of muons in the mixed perovskite lattice.
     
%
%
%
%
%
%
%
%
%
%

\section{Materials and methods}\label{sec2}
LSAT single crystals of 10$\times$10$\times$0.5~mm$^3$ were obtained
from CRYSTAL GmbH, Germany.
These single crystals were grown by the Czochralski method and cut along
the cubic (001) plane.
Time-differential $\mu^+$SR experiments were performed at J-PARC using a
spin-polarized surface muon beam in a longitudinal field configuration, 
with its initial polarization direction nominally parallel to the beam axis.
 The ARTEMIS spectrometer~\cite{kojima18} at the S1 area of the 
 Materials and Life Science Experimental Facility was used
 with a conventional $^4$He flow cryostat for the ZF-$\mu^+$SR measurements
 of LSAT in the temperature range from 4~K to 270~K. Four single
 crystals of LSAT were mounted on a silver sample holder with the (001)
 plane perpendicular to the beam axis.
   The asymmetry of $\mu^+$ decay, $A$, was monitored by the ``Forward''
 and ``Backward'' positron counters as a function of the 
 elapsed time $t$ from the instant of muon implantation.

All DFT calculations were performed using the QUANTUM ESPRESSO
package~\cite{giannozzi09,giannozzi17}.
The generalized gradient approximation using Perdew-Burke-Ernzerhof
(PBE) exchange correlation functional was adopted. Projector
augmented-wave potentials with La(4$f$, 5$s$, 5$p$, 5$d$, 6$s$, 6$p$),
Sr(4$s$, 4$p$, 5$s$, 5$p$), Al(3$s$, 3$p$), Ta(5$s$, 5$p$, 5$d$, 6$s$)
and O(2$s$, 2$p$) valence states were used. Hubbard $U$ corrections were
applied to the La 4$f$ and Ta 5$d$ states using typical $U_{eff}$ or
$U-J$ values listed in Ref.~\cite{calderon15}.
Electron wave functions were expanded in plane waves
with a cutoff of 70 Ry for kinetic energy and 600 Ry for charge density.

The LSAT structure was simulated within a $3\times3\times3$ supercell
of the five-atom cubic primitive cell. The supercell was composed of 7 La,
20 Sr, 17 Al, 10 Ta, and 81 O atoms so that it maintains total charge
neutrality and approximates the chemical composition of our sample
(La$_{0.26}$Sr$_{0.76}$Al$_{0.61}$Ta$_{0.37}$O$_3$ according to the
crystal manufacture-supplied data sheet~\cite{lsat_data}) well.
To simulate the random cation distribution in LSAT, a special
quasi-random structure (SQS) was generated using the Alloy Theoretic
Automated Toolkit (ATAT)~\cite{walle08} and applied to the supercell. 

Brillouin zone sampling with a Monkhorst-Pack {\bf \textit{k}}-point
mesh of $2\times2\times2$ and Gaussian smearing with a broadening of 0.01 Ry
were applied for structure optimization calculations using the pw.x
module in the QUANTUM ESPRESSO package.
Atomic positions were relaxed, while the lattice constant for the
supercell was kept fixed at 11.604\AA , according to the experimental
lattice constant reported in the crystal manufacture-supplied data
sheet~\cite{lsat_data}.
Atomic forces were converged to within $7.7\times10^{-3}$ eV/\AA~ in the
optimization process.

The spatial distribution of electrostatic potential in the optimized
supercell was calculated using the pp.x module. It should be noted that
the supercell does not involve any muon equivalent, and therefore the
following discussion about the stability of muons at lattice interstices
is simply based on a local electric field approximation.

We also evaluated the spatial distribution of
 $\gamma_{\mu}\sigma$ in the optimized supercell via a dipolar
 sum calculation, where $\gamma_{\mu}$ is the muon gyromagnetic ratio 
 and $\sigma$ is the root-mean-square width of the nuclear dipole field
 distribution at a given point.
 Through this calculation, both the experimental geometry using the
 single-crystalline sample and the effect of $\mu^+$-induced electric
  quadrupole interactions on nearest neighbor nuclei at cation sites were
 taken into account~\cite{hayano79}.

\section{Results and discussion}\label{sec3}

\begin{figure}[h]%
\centering
\includegraphics[width=1.0\textwidth]{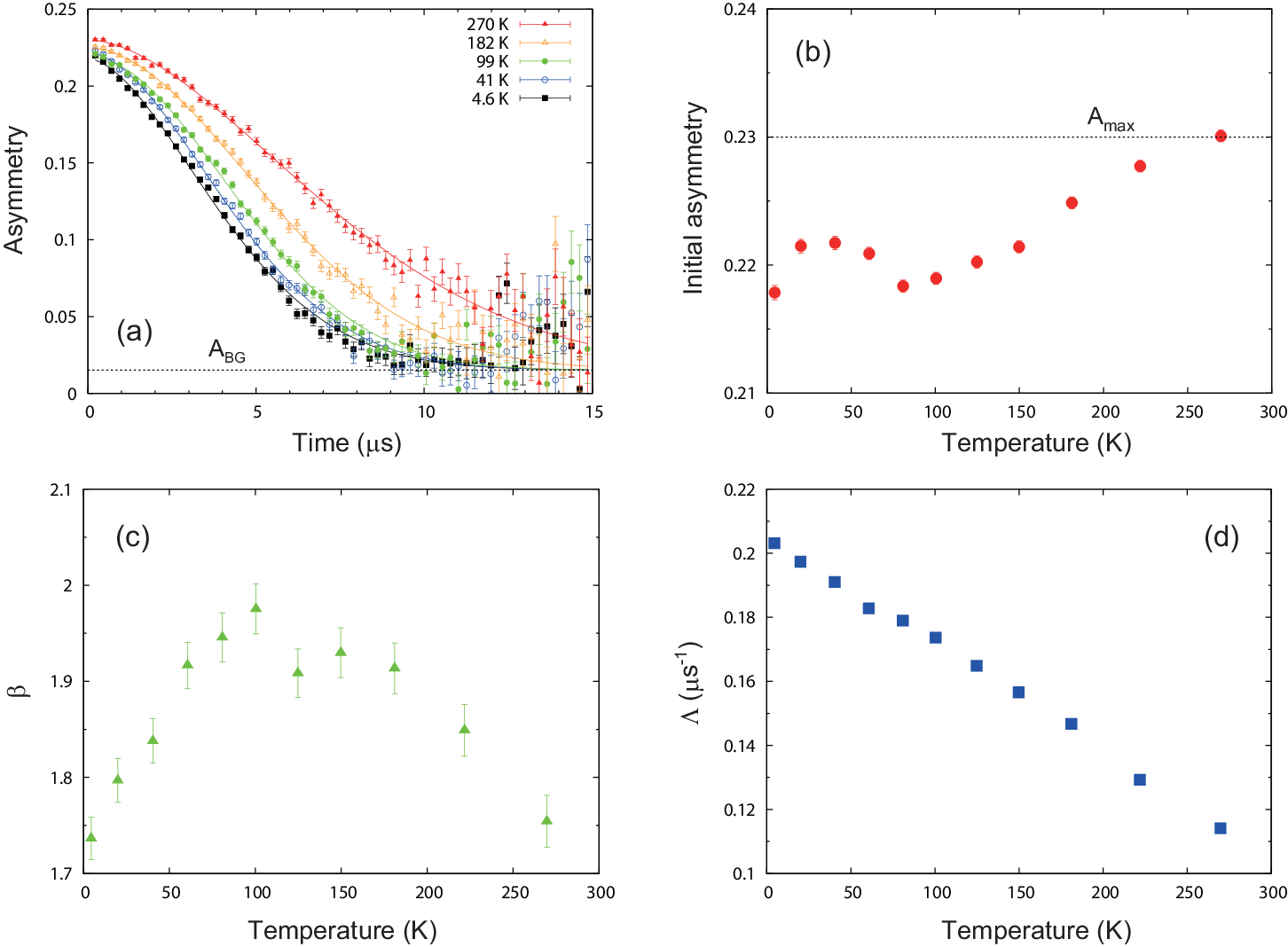}
\caption{(a) ZF-$\mu^+$SR spectra at 4.6,
 41, 99, 182, and 270~K. (b) to (d) Temperature dependencies of initial
 asymmetry ($A_{\rm s}+A_{\rm BG}$), $\beta$, and $\Lambda$.}\label{fig1}
\end{figure}

Figure \ref{fig1}(a) shows ZF-$\mu^+$SR spectra, $A(t)$, at 4.6,
41, 99, 182, and 270~K, representing the depolarization of implanted muons
through interactions with surrounding nuclear spins.
To capture the characteristics of the temperature variation, the
following function was used to parameterize the spectra, i.e.,
\begin{equation}
A(t) = A_{\rm s} \exp(-(\Lambda t)^{\beta}) + A_{\rm BG},\label{eq1}
\end{equation}
where $A_{\rm s}$ and $A_{\rm BG}$ are partial asymmetries from muons
stopped in the sample and the silver sample holder, respectively,
$\Lambda$ is the depolarization rate, and $\beta$ is the exponent.
The magnitude of $A_{\rm BG}$ was determined using a reference sample
(holmium) with similar geometry to the LSAT sample.
The temperature dependencies of initial asymmetry ($A_{\rm s}+A_{\rm
BG}$), $\beta$, and $\Lambda$ are shown in Figs.~\ref{fig1}(b) to
(d).

The initial asymmetry is close to the device-specific maximum value 
$A_{max}$ of 0.23, suggesting that muons implanted into LSAT
are predominantly found in diamagnetic environments.
A small missing fraction at low temperatures may be associated with
metastable paramagnetic species formed upon 
high energy implantation of muons, which were also observed in
other transition-metal oxide insulators~\cite{ito13,salman14,vieira16,ito19,ito22}.
In this paper, we focus on the main diamagnetic component and do not
elaborate further on the small paramagnetic fraction.

The Gaussian-like line
shape observed at 4.6~K suggests that muons are quasi-static in the
$\mu^+$SR time window ($\sim 10^{-5}$~s) at low temperatures.
Interestingly, the Gaussian-like feature is maintained up to room
temperature (Fig.~\ref{fig1}(c)), while the depolarization rate
monotonically decreases with increasing temperature (Fig.~\ref{fig1}(d)),
suggesting the onset of muon diffusion.
This is in sharp contrast to the 
depolarization behavior that is characteristic of global muon diffusion in
homogeneous dense nuclear spin systems, as described by the dynamical Gaussian
Kubo-Toyabe function~\cite{hayano79,storchak98,ito23}.
The motional narrowing that preserves the Gaussian-like line shape 
can occur when the autocorrelation function of the nuclear dipole field
experienced by muons has a long-time correlation term~\cite{ito24}.
Such a situation can be realized not only when muons simultaneously
observe diffusing and static ions with non-zero nuclear dipole moments
at fixed sites~\cite{ito24}, but also when each muon locally jumps
between a few adjacent sites in a confined space of the angstrom
scale.
The local hopping picture may be justified in the LSAT lattice because 
the random distribution of four cations with different nominal valences 
can cause a large modulation in electrostatic potential at the
electron-volt scale, which significantly limits the space where muons
can diffuse.

\begin{figure}[h]%
\centering
\includegraphics[width=1.0\textwidth]{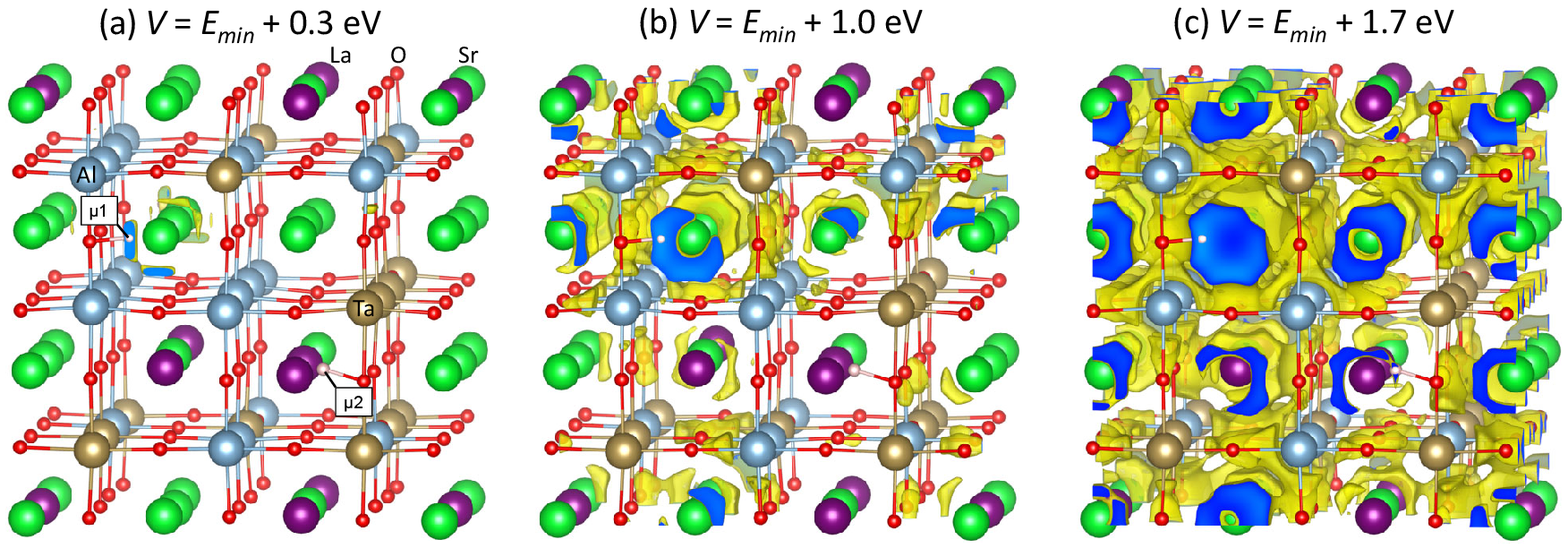}
\caption{Isosurfaces for electrostatic potential $V$ in the
 3$\times$3$\times$3 SQS supercell of LSAT at (a) $E_{min} + 0.3$~eV,
 (b) $E_{min} + 1.0$~eV, and (c) $E_{min} + 1.7$~eV, drawn with yellow curved
 surfaces with bluish cross sections. The smallest spheres labeled 
 $\mu 1$ and $\mu 2$ exhibit the minimum energy position at $E_{min} $ and a
 representative high energy position at $E_{min} + 1.6$~eV, respectively.}\label{fig2}
\end{figure}

The electrostatic potential $V$, calculated for the SQS model of LSAT, further
justifies the local hopping picture.
Figure~\ref{fig2} exhibits the isosurfaces at (a) $V=E_{min} + 0.3$~eV,
 (b) $V=E_{min} + 1.0$~eV, and (c) $V=E_{min} + 1.7$~eV in the
 3$\times$3$\times$3 SQS supercell, where $E_{min}$ is the global
 minimum energy.
 Local minima in $V$ are found at interstitial positions approximately
1\AA~away from nearest neighbor O in $B$O$_2$ planes, which are
consistent with $\mu^+$ sites identified in other perovskite
oxides~\cite{hempelmann98,ito17,ito19,ito22,ito23}.
The $A$ and $B$ cation sites around the minimum energy position ($\mu 1$)
with $V=E_{min}$ are mainly filled with Sr$^{2+}$ and Al$^{3+}$, respectively,
whereas $V$ is even higher by 1.6~eV at a representative oxygen-bound position
($\mu 2$) surrounded by La$^{3+}$ and Ta$^{5+}$ cations.
This is perfectly in line with the intuition that positively charged
muons tend to avoid interstitial positions close to La$^{3+}$ at the $A$
site or Ta$^{5+}$ at the $B$ site with relatively large nominal valences.

The evolution of the volume enclosed by the isosurfaces of $V$ from
Fig.~\ref{fig2}(a) to (c) suggests that muon diffusion paths
are significantly restricted and isolated below room temperature, although
they could expand with further increasing temperature through their interconnection.
This implies that global muon diffusion is substantially inactive below
300~K in LSAT, in sharp contrast to the situation in a homogeneous
KTaO$_3$ lattice, where global diffusion was observed even below room
temperature~\cite{ito23}.

\begin{figure}[h]%
\centering
\includegraphics[width=0.5\textwidth]{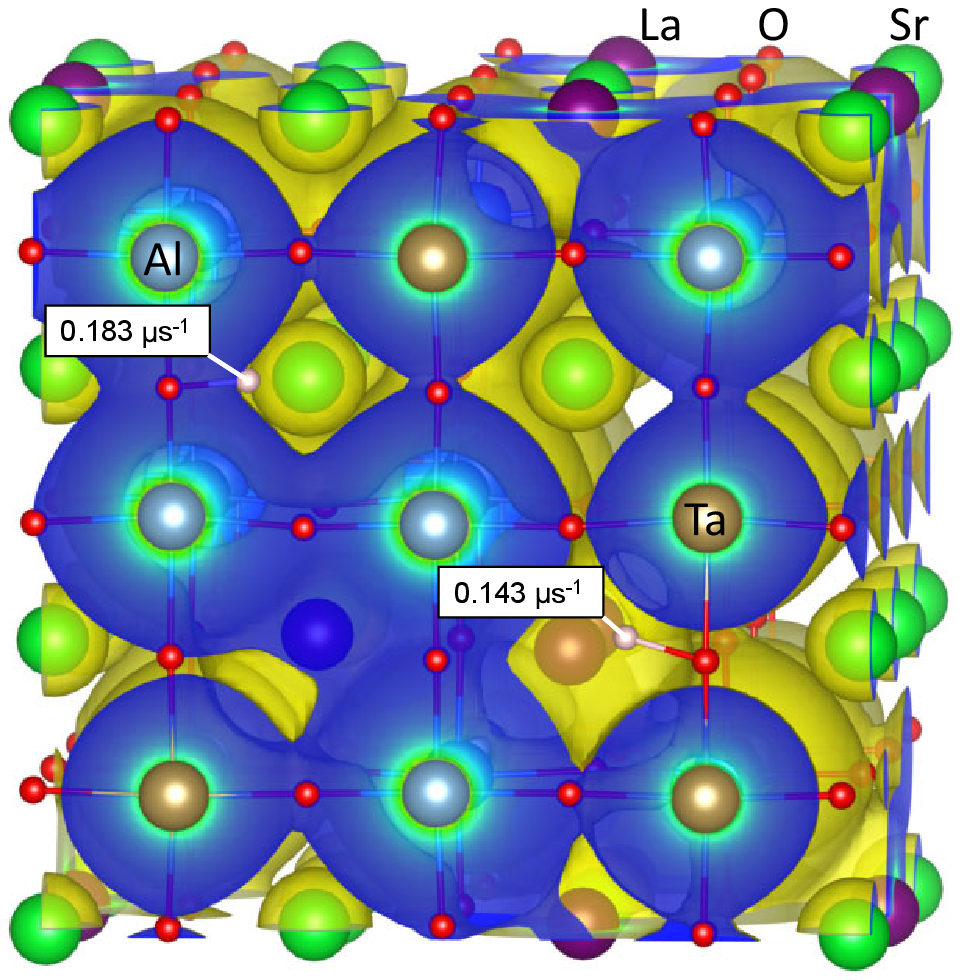}
\caption{Isosurface for $\gamma_{\mu}\sigma=0.20~\mu$s$^{-1}$ in the
 3$\times$3$\times$3 SQS supercell of LSAT, drawn with yellow curved
 surfaces with bluish cross sections.}\label{fig3}
\end{figure}

Finally, we discuss the consistency between the static line width
$\gamma_{\mu}\sigma$ calculated for the SQS model and the experimental
value of $\Lambda$.
Figure~\ref{fig3} shows the isosurface for
$\gamma_{\mu}\sigma=0.20~\mu$s$^{-1}$ in the 3$\times$3$\times$3 SQS
supercell of LSAT, which corresponds to $\Lambda$ at
$T\rightarrow 0$~K (Fig.~\ref{fig1}(d)). Oxygen-bound positions
surrounded by Sr$^{2+}$ and Al$^{3+}$, which are represented by the
$\mu$1 site with $\gamma_{\mu}\sigma = 0.183~\mu$s$^{-1}$ and $V=E_{min}$,
are close to the $0.20$-$\mu$s$^{-1}$ isosurface. On the other hand, those 
surrounded by La$^{3+}$ and Ta$^{5+}$ with much higher $V$ are slightly off the
isosurface, as evidenced by the smaller value of
$\gamma_{\mu}\sigma=0.143~\mu$s$^{-1}$ at the $\mu$2 site.
Since $V$ at the $\mu$2 site is 1.6~eV higher than that at the $\mu$1
site, it is unlikely that the decrease in $\Lambda$ toward
room temperature is mainly caused by an increase in the relative
occupancy of such high $V$ sites with smaller $\gamma_{\mu}\sigma$s.
It seems more plausible that the thermal activation of the local muon
hopping between a few oxygen-bound sites surrounded by Sr$^{2+}$ and
Al$^{3+}$, such as the $\mu$1 site, would cause the decrease in
$\Lambda$.

\section{Conclusions}\label{sec4}
In summary, we performed ZF-$\mu^+$SR measurements on LSAT
mixed-perovskite single crystals. Muon depolarization due to
the distribution of nuclear dipole fields was observed in the
temperature range from 4~K to 270~K. While the $\mu^+$SR time spectrum
maintained a Gaussian-like feature over the entire range, the
depolarization rate $\Lambda$ exhibited a monotonic decrease with increasing 
temperature. This behavior may be attributed to the thermally
activated diffusion of muons between a few adjacent sites within a 
confined space of the angstrom scale (local hopping), where the
motionally averaged local field that each muon feels can remain non-zero 
and result in maintaining the Gaussian-like line shape.
The spatial distribution of electrostatic potential $V$ at lattice
interstices evaluated via DFT calculations suggests
that such a restriction of muon diffusion paths can be caused by the
random distribution of cations with different nominal valences in the
mixed perovskite lattice.

In terms of applications, LSAT is widely used as a single-crystalline substrate
for the epitaxial growth of thin films.
In $\mu^+$SR investigations of thin films deposited on the
LSAT substrate using ultra-slow muons at J-PARC or low-energy muons at
the Paul Scherrer Institut, attention must be paid to the contamination
of the temperature-dependent signal from the substrate.

\backmatter

\bmhead{Acknowledgments}
We thank the staff of the J-PARC muon facility for technical
 assistance. The DFT calculations were conducted with the supercomputer HPE
SGI8600 in Japan Atomic Energy Agency. This research was partially
 supported by Grants-in-Aid (Nos. 23K11707, 21H05102, 20K12484, 20H01864, and
 20H02037) from the Japan Society for the Promotion of Science.

\bmhead{Author contributions}
T.U.I. conceived the study, performed the $\mu^+$SR experiment in
collaboration with W.H., A.K., J.G.N., and K.S., conducted data analysis
and DFT calculations, and wrote the manuscript.
All authors participated in the review and editing of the manuscript.




\end{document}